\documentclass{elsart}

\usepackage{graphicx}

\begin{document}

\title{Stock Market Scale by Artificial Insymmetrised  Patterns}
\author{D.~Makowiec}
\address{Institute of Theoretical Physics and Astrophysics, 
Gda\'nsk University, 80-952~Gda\'nsk, ul.Wita Stwosza 57, Poland,
fizdm@univ.gda.pl}

\maketitle

{\bf Abstract}\\
{\small 
Large and stable indices of the world wide stock markets such as NYSE and S$\&$P500 together with  NASDAQ -- the index representing markets of new trends, and WIG -- the index of the local stock market of Eastern Europe, are considered.  Due to the relation between  artificial insymmetrised patterns (AIP) and time series, stationary and temporary properties of stock market indices are identified. By filtering extreme events it is found that fluctuations  are self-similar. Snap-shots in time lead to estimates for a temporary state of a market with respect to its history. It appears that close to a  crash the AIP representation of a system  becomes frozen.
}

Keywords: econophysics, empirical study,  visualization of data 

PACS code :  05.40-a 89.90+n

\baselineskip = 24pt

\section{Introduction}
It is not disputed that developing ways to measure the state of financial market accurately is essential, see, e.g., \cite{Olsens}. The purpose of this paper is to show how using a simple  graphical tool we can estimate a state of an asset. The basic idea comes from the use of visual displays. An Artificial Insymmetrised Pattern, AIP in short, was introduced by Pickover  in 1986 \cite{Pickover1} to describe human voice and animal vocalization. The method, called also Symmetrised Dot Patterns, takes advantage of the fact that symmetry, color and redundancy of dot patterns are useful in the visual detection and memorization of patterns by any human analyst. The method was successfully applied to many different time series \cite{Pickover2}, becoming a technique for the qualitative assessment of time series \cite{Posiewnik}. With AIP, one can graphically detect hidden patterns and structural changes in data or see similarities in patterns across the time series being studied.

The fundamental assumption underlying the idea of AIP or any similar type propositions, like, e.g. recurrence plots \cite{EckmanRuelle}, or $i$-variability diagram method \cite{Ausloos}, is that an observable time series is the representation of some dynamical process. 
Typically, we have interactions between many different factors over time which finally sum up into a one dimensional time series. An example of such process is a stock market.  The behavior of the stock market is determined by many factors, such as the economic and political environment, investors expectations and individual trader's decisions. But eventually we deal with a single factor, namely, the price of a stock. 

In the following we propose how to read AIPs if  time series of indices of different stock markets are studied. Properties of NYSE Composite (NYSE in short) and Standard and Poors 500 (S$\&$P 500 in short) lead us to a picture of a well developed  and mature market. A young market, Warsaw Stock Exchange, which just celebrated its 11th anniversary this April, is  a representative of so-called emerging markets. Its index is called WIG. We also look at NASDAQ Composite (NASDAQ in short) to take inspect a market of a high volatility  related to the "Internet Economy". Some AIPs for the listed series are presented in \cite{beauty,Makowiec}. With this paper we maintain that such stylized facts known for fluctuations of financial time series as (a) weak correlations between the price changes on succesive trading days, (b) positive kurtosis like in a Levy noise and decay with a power law in tails for the probability distribution function of the price changes, and (c) correlations between the absolute values of the price changes, called clustering of volatility, are easily readable from dot patters, Section 3. Moreover, by extracting extreme events we   find that fluctuations could be self-similar  and if they are not self-semilar we obtain the background noise that drives the market, Section 4. Finally, by snap-shots in time we show how to estimate a state of a market with respect to its history. It appears that close to a  crash, the system becomes frozen - the dependence on subsequent points is fixed, Section 5.

\section{AIP method}
The procedure is to convert a series of data $ \{ x(t) \}$ into a collection of dots by a simple computer program: 
\begin{itemize}
\item[---] choose the model parameters  $R$ and  $\tau$  suitable to your needs,
\item[---] rescale the data to the range $[0,R]$ using the formula: 
$$ x(t) \leftarrow R \quad {x(t)-x_{min}\over x_{max} - x_{min} } $$
where $x_{min} =\min_{t}\{ x(t) \}$ and $x_{max} =\max_{t}\{ x(t) \}$ 

Thus the low values of a series are close to  $0$ and high values are near $R$. One can say that each series is presented in its own min-max scale. 
\item[---] mark  a dot on a 2-dimensional plane using subsequent points  separated by $\tau$  but in  polar coordinates {\it(radius, angle)}:
$$ \{  ( x(t), x(t+ \tau) )\}$$
\item[---] multiply a plot symmetrically   by rotating $k$ times by $\phi$ angle
$$ \{   ( x(t), x(t+ \tau) + k\phi )\}$$
and/or  reflecting trough mirror planes :
$$ \{  ( x(t), \pm[ x(t+ \tau) + k\phi ]) \}$$
\end{itemize}
The resulting representation of a series, called Artificial Insymmetrised Pattern, provides a base in which local visual correlation can be used in detection and characterization of significant features of any sample data.
The gallery of different AIPs can be found in \cite{beauty,Makowiec}. 

Our present interest is to compare different series  or different epochs in a series and therefore in all figures included here we show patterns where most of symmetries are left out --- we plot different series on the same figure to observe lack of symmetries. For the best presentation of data we choose $ R$ to be equal to $\pi$. As a result, a single dot pattern is plotted on a half of a circle of radius $\pi$.  With the rescaling $ R=\pi$ the  plots of standard series take the form presented in Fig.1.  The basic set of standards consists of  noises: Gaussian, white and Levy at  different  $\alpha$s ($\alpha $ represents decay of tails) and  a random walk. Let us emphasize that all noises do not give different patterns when the distance between subsequent points  $\tau$ is changed. The random walk depends on $\tau$ - with increasing $\tau$ the width of a curve increases. We also include a picture of a walk made from a half of points of  a series to show regularity in density of a resulting curve.

Having in mind the visual scale just shown, let us move to estimates of states of stock markets.

\section{Stylized facts by AIP}

The basic quantities studied here are values of indices: WIG, NASDAQ, NYSE and S\&P500 at closing. For each time series we consider a series of daily returns defined as  usual \cite{Gopik}:
\begin{equation}
{\rm return}(t+1) = \ln[ {\rm index}(t+1)]- \ln [ {\rm index}(t)]
\end{equation}
The analyzed data comes from time span between 1st of January 1995 and 31st of December 2001 (7-year study resulting in about 1750 data points) or between 1st of January 1999 and 31 December 2001 (3-year study with about 750 data points). 

The min-max scale of AIP  is determined by extreme events. These events are located close to  the circle of $\pi$  radius - positive tails, or at the zero - negative tails. Fig. 2 is a density AIP plot to compare returns of last seven years to returns of last three years. All these patterns are peaked significantly more than a Gaussian noise. 
Presence of rare events squeezes the "body" of AIPs  making the area of the AIP plot much smaller than if the AIP plot is of a Gaussian noise. Comparing plots of Fig. 2 to the scale plots of Fig. 1 we see that plots of returns  are similar to the Levy noise with $\alpha $  about $ 1.8 $ in case of 7-year data and $\alpha  > 1.9 $ in case of 3-year data. The tendency of flattening of distributions can be easily verified by calculating kurtosis. It appears that the kurtosis for 7-year data is: WIG: $ 2.607$, NASDAQ: $4.302$, NYSE: $4.652$ and S$\&$P 500: $3.710$ and if 3-year data is considered then WIG: $ 1.878$, NASDAQ: $1.667$, NYSE: $1.761$ and S$\&$P 500: $1.261$. Moreover, the localization of bodies of the 7-year  data is moved from the vertical axis to the left in case of NYSE, S$\&$P 500 and WIG and to the right when NASDAQ is considered. It is because in a scale of individual min-max events at average the returns occupy rather: higher half of values (NYSE, S$\&$P 500, WIG) or lower half (NASDAQ).

Rare events lead to heavy tails in the distribution of returns \cite{old,first}. What we can read from Fig. 3 is that the area where positive rare events are plotted is more densely populated in case of indices than in case when any Levy noise is considered. Therefore, we can speculate that tails of the distribution of indices decay diferently from a Levy noise world. 

It is interesting to compare walks performed by logarithms of indices, see Fig. 4.  The 7-year data, Fig. 4(a), shows a walk which in case of NYSE and S$\&$P 500 is pushed to high values. NASDAQ's walk is rather located close to the zero. WIG's walk  shows a similar feature. The 3-year data, Fig. 4(b), shows as all walks have been shifted towards low values. Moreover, the events separated by 20 days are scattered over a half of a circle more widely than it is with 7-year walk. Thus the considered stock walks when observed in last three years are definitely more noisy than if they are considered from the perspective of 7 years.

It is generally assumed that for probability distribution function of volatility a log-normal distribution is a good approximation \cite{Olsens}.
In Fig. 5 we present AIPs for volatility time series.  It concludes that one-day volatility(: absolute values of returns) Fig. 5(a),  for three of four indices is rather far from being a Gaussian noise, while  five-day volatilities, especially, when  $\tau =20$, Fig. 5(b), appear to be like a Gaussian noise. The five-day volatility is calculated as standard deviation of five succesive  returns.

To observe clustering of volatility we divide series into epochs and we study  where points of a considered epoch are located in a main body of volatility dot patterns (see subsequent plots of  Fig. 6.)  In particular, Fig. 6(a) shows the whole 3-year data of one-day volatility and 5-day volatility. Fig. 6(b) presents volatility in the first three months of 1999. Notice, that volatility of Warsaw market is high what is different from volatility of the  New York markets. The next snap-shots are taken at the moment where the opposite situation happens: a high volatility  term  is going on  New York markets while Warsaw market experiences a calm time,  Fig. 6(c). Fig. 6(d) shows data from a month before the attack of 11 September 2001  and Fig. 6(e) - a month after the attack. Notice, that some indices were not influenced by this event. Finally, the present state of markets is shown on Fig. 6(f) where the data   from the last month of a period studied is presented. All markets are quiet.

\section{Filtering data}

As  the main shape of an AIP is definitely generated by extreme events, we should ask what kind of a pattern would arise if the most exotic  points were extracted from a series. On the following figures, Figs 7-10,  we show results  when subsequently a hundred, two-hundred and four-hundred points are extracted from the series. 

In case of NYSE walk, Fig. 7(a), just after filtering a hundred points, the data separated by 20 days starts to appear as a white noise rather than a walk. However, the  filtered returns, Fig. 7(b),  ensures us that the noise is still non-Gaussian and the peaked distribution is firmly kept. We interpret this feature as a self-similarity. The series of S$\&$P 500, Fig. 8(a), shows also that elimination of  extreme events makes a series more noisy. Returns, Fig. 8(b), when filtered subsequently, slightly move the center of a body what suggests that the max values are distant from the average value. NASDAQ walk data when filtered, Fig. 9(a), shows distinct properties.  The area of low values stays dense occupied and, although more and more points are extracted, the walk arrives to high values rarely. The peaked distribution of returns is kept, however, and we observe a move of the main body down, Fig. 9(b). The walk of WIG, Fig. 10(a), seems to be in a state between NYSE and NASDAQ --- the dots of a plot rather quickly are scattered over the whole  circle. However the distribution of dots is not uniform. Low values are occupied more densely. Surprisingly, the returns of WIG, Fig. 10(b), after filtering loose the positive kurtosis and become similar to a Gaussian noise.

\section{Crashes}
By study larger values of $\tau$ like $\tau= 50, 100$ we can observe how the system arrives to its minimal and maximal values, see Fig. 11. Notice that WIG and NASDAQ maxima are separated dots, minima are densely occupied. 
Moreover, these walks are squeezed along the horizontal axis. Hence large values are followed by strong drops. This feature does not appear for plots of  NYSE and S$\&$P 500. We know that during the period we studied NASDAQ experienced the crash. It was in April of 2000. At the same time  huge drops were noticed on Warsaw stock market.  Is it possible that this event can be recognized by AIPs? Can we learn in advance about an approaching crash from dot patterns? 

Fig. 12 presents plots for series of data taken before the crashes. NASDAQ and WIG plots  consist of  data of two years before April 2000 (March values are the last data included). For NYSE and S$\&$P 500 we consider data of  two years  before October 87 crash (September is the last data included). 
Although data on Fig. 12(a) and Fig. 12(b) are plotted with increasing delay between points, namely $\tau=10,20,50,100$, the resulting plots look both differently and similarly. They are similar to each other  because there is no visible influence of $\tau$ (despite squeezing the body). Markets look like being frozen --- driven by some deterministic rule. Moreover, the points of the top of walks are distributed loosely while the lowest values are densely occupied. Plotting snap-shots of last 60 days,  we see that in the last days there are flights rather not walks to high values, see Fig. 13.

In the period that we studied  one more crash took place. At the end of August 1998, S$\&$P 500 lost $6.79\% $. That crash was different from others discussed so far, because is was a crash of confidence. It was caused by events in Asia and Russia. Again, when plotting AIPs from data before the crash we see that the markets are frozen, Fig 14. The Warsaw market behaves very noisy at that time. It looks like WIG experienced a crash earlier.
\section{Summary}
We found that visualization of the economic data by the AIP method is  particularly useful in getting into relations between markets. 
In particular: \\
(1) The  stylized facts of returns of financial time series: lack of correlations in daily returns, positive kurtosis of probabability density function of returns  with tails that are different from both Gaussian and any Levy noise, and clustering of volatility are easily readable from AIPs. By differentiating  these features we can compare markets. \\
(2) Filtering out the extreme events we can learn about fractal properties of the data or un-hide  the background noise that drives fluctuations. \\
(3) Moreover, as the AIP method keeps points in the  time order, we can also easily find nonstationary features of markets. By taking snap-shots in time we can estimate a market state with respect to its history. Especially this property seems to be important in recognizing the ill condition of a market such as an imminent crash.

Wealth of observations together with simplicity of use make the graphical method presented in this paper worth to be taken into account.
For example, due to observations that are derived only because of using AIPs, we are not surprised by the opinion that eastern Europe markets are driven by different logic than other markets, \cite{Johansen-Sornette}.

{\bf Acknowledgement\\}
This work is supported by  Gda\'nsk University: BW 5400-5-0235-2

\newpage
\begin{figure}
\caption{The visual (eye) AIP scale of basic noises. For further analysis it is useful to remember localization and shape of patterns. Notice the dependence on $\tau$ in case of a random walk and that a shape of a walk is independent of on number of points used. }
\end{figure}

\begin{figure}
\caption{Density plots of AIPs for series of returns of stock indices. 7-year data --  upper half of a circle, is plotted against to  3-year data -- lower half of a circle. Notice the change in  sharpness of tops of distributions of points when we move from 7-year data to 3-year data. }
\end{figure}

\begin{figure}
\caption{AIPs for returns of: WIG, NASDAQ, NYSE and S$\&$P 500 for data of  7 (black dots) and 3 (gray dots) years versus a cicrle of  extreeme events.}
\end{figure}

\begin{figure}
\caption{AIP plots for logarithms of subsequent indices.  Data comes from (a) 7 years, (b) 3 years.   Black plots represent one-day delay, $\tau=1$, gray dots are plots with 20 days delay, $\tau=20$. }
\end{figure}

\begin{figure}
\caption{AIPs for logarithms of (a)1-day (b) 5-day  volatility. Black plots represent one day delay, $\tau=1$, gray dots are plots with  20 days delay, $\tau=20$. }
\end{figure}

\begin{figure}
\caption{ Study clustering of volatility. 
(a) AIPs for the whole data of one-day (black dots) and five-day (gray dots) volatility;
(b) first three months of 1999 on markets. Notice high volatility of WIG;
(c) a period  of high volatility of New York indices:  17 April-17 May 2001;
(d) a month before the September attact on New York;
(e) a month after the September attact on New York;
(f) last month, December 2001, on studied markets. 
}
\end{figure}

\begin{figure}
\caption{Extracting extreme events. (a) AIPs for a walk of NYSE when maximal and minimal points are subsequently removed (b) AIPs for returns of NYSE.}
\end{figure}

\begin{figure}
\caption{Extracting extreme events. (a) AIPs for a walk of S$\&$P500 when maximal and minimal points are subsequently removed (b) AIPs for returns of 
S$\&$P500.}
\end{figure}

\begin{figure}
\caption{Extracting extreme events. (a) AIPs for a walk of NASDAQ when maximal and minimal points are subsequently removed (b) AIPs for returns of NASDAQ.}
\end{figure}

\begin{figure}
\caption{Extracting extreme events. (a) AIPs for a walk of WIG when maximal and minimal points are subsequently removed (b) AIPs for returns of WIG.}
\end{figure}

\begin{figure}
\caption{ AIPs for indices with increasing delay between points (a) $\tau=50 $ (b) $\tau =100$.}
\end{figure}

\begin{figure}
\caption{Two years before crashes: (a) $\tau=20 $ versus $\tau =10$ (b) $\tau=50 $ versus $\tau =100$. Notice similarities between all patterns for a given market and between different markets.
}
\end{figure}

\begin{figure}
\caption{Last days before crashes. Notice that prices are far from their average historical values. }
\end{figure}

\begin{figure}
\caption{August 1998 crash --- AIPs for two-year data before the crash: (a) $\tau=20 $ versus $\tau =10$ (b) $\tau=50 $ versus $\tau =100$.  }
\end{figure}

\end{document}